\documentclass[aps,amsfonts,amsmath,prd,preprint,nofootinbib]{revtex4}
\newcommand{\beq}{\begin{equation}}
\newcommand{\eeq}{\end{equation}}

\usepackage{epsfig}

\begin{document}

\title{Quantum Tunneling in Flux Compactifications}

\author{Jose J. Blanco-Pillado}
\author{Delia Schwartz-Perlov}
\author{Alexander Vilenkin}

\affiliation{Institute of Cosmology, Department of Physics and Astronomy,\\ 
Tufts University, Medford, MA 02155, USA}

\def\changenote#1{\footnote{\bf #1}}

\begin{abstract}
We identify instantons representing vacuum decay in a
6-dimensional toy model for string theory flux compactifications,
with the two extra dimensions compactified on a sphere.  We evaluate
the instanton action for tunneling between different flux vacua, as
well as for the decompactification decay channel.  The bubbles
resulting from flux tunneling have an unusual structure.  They are
bounded by two-dimensional branes, which are localized in the
extra dimensions.  This has important implications for bubble
collisions.
\end{abstract}

\maketitle
\section{Introduction}

String theory suggests the existence of a multitude of vacua
characterized by different values of the low-energy constants of
Nature \cite{Bousso:2000xa,KKLT, Susskind}.  When combined with
inflationary cosmology, this leads to the picture of an eternally
inflating ``multiverse'', populated by all possible types of vacua.
Transitions between different vacua occur through nucleation of
bubbles and their subsequent expansion.  The calculation of bubble
nucleation rates is therefore one of the key problems one needs to
solve for a quantitative description of the multiverse.

More recent developments, related to the so-called Boltzmann brain
paradox, make this problem especially acute.  Boltzmann brains are
``freak'' observers who spontaneously pop out in the vacuum as a
result of quantum fluctuations.  Even though their formation rate is
extremely small, they may greatly outnumber regular observers, unless
the vacuum decay rate is sufficiently high in all the vacua that can
support Boltzmann brains \cite{BB1,BB2}.  This imposes an unexpected and
somewhat restrictive constraint on possible vacuum decay rates in the
string theory landscape.  Some recent work suggests that this
constraint may indeed be satisfied for KKLT-type vacua \cite{FL}, but
the issue is far from being settled.

In a four-dimensional field theory, different vacua correspond to
minima of a scalar field potential, separated by barriers.  A
formalism for calculating the bubble nucleation rate in this
framework has been developed in the classic paper by Coleman and De
Luccia \cite{CdL}.  String theory introduces a number of
complications.  We have to deal with a higher-dimensional spacetime,
in which the extra dimensions are compactified.  The role of scalar
fields is played by the moduli that characterize the sizes and other
geometric aspects of these extra dimensions.  String theory vacua
also involve additional objects, such as fluxes and branes.  Bubble
nucleation rates in semi-realistic superstring vacua have been
studied in the literature; see, e.g., \cite{Kachru-tunel,Frey,FL,Lust} for
recent discussion and references.

Our goal in this paper is to study the bubble nucleation rate in a toy
model of the landscape, which is rich enough to include some of the
essential features of the ``real thing'' and at the same time simple
enough to allow a quantitative analysis.  As a warm-up exercise, we
shall first consider vacuum decay in some lower-dimensional models
(Sections II and III), but our main focus will be on a 6-dimensional
Einstein-Maxwell theory, with the extra dimensions compactified into a
2-sphere and their radius stabilized by a magnetic flux through that
sphere.  This model has a long pedigree \cite{Freund-Rubin,EM6D}; more
recently it has been discussed as a toy model for string theory
compactification \cite{flux-compactifications}.  We shall show that
vacuum decay in this model can occur through the nucleation of
magnetically charged 2-branes, which look like expanding spherical
bubbles in the large 3 dimensions and are localized in the extra 2
dimensions.  The vacuum inside the bubble has its extra-dimensional
magnetic flux reduced by one unit compared to the vacuum outside.  We
shall estimate the corresponding instanton action and compare it with
that for the alternative channel of vacuum decay -- the
decompactification of the extra dimensions.
Finally, we shall discuss some unusual properties
of flux vacuum bubbles, in particular with regard to bubble
collisions.

\section{Lower-dimensional examples}
\subsection{$(1+1)$ dimensions}
Perhaps the simplest model that we can use to visualize the
type of process that we are interested in is a $1+1$
dimensional spacetime, where the spatial dimension
is compactified. Let us consider a Lagrangian of the 
form
\beq
S_{1+1}=\int{dt~dy \left(-  {1\over 2} \partial_{a} \phi
  {\partial^{a} {\bar \phi}} - {{\lambda}\over {4}} (\phi {\bar \phi} - \eta^2)^2\right)},
\label{2D-action}
\eeq
where $a=y,t$ are the two dimensions in this toy model and
we are assuming that the spatial dimension is 
compact, $0 < y < L$. The equation of motion for this
model is
\beq
\partial_a \partial^a \phi =\lambda \phi (\phi {\bar \phi} - \eta^2)~.
\label{complex-scalar-eom}
\eeq
We can look for a static solution to this equation by 
assuming that the complex scalar field winds $n$ times around the
compact dimension, namely,
\beq
\phi(y) = {\tilde \eta} e^{i \theta(y)} ={\tilde \eta} e^{ i {{2 \pi n y}\over L}}~,
\label{phiy}
\eeq
which is a solution of the previous equation of motion provided
that
\beq
{\tilde \eta}^2 = \eta^2- {{4 \pi^2 n^2}\over {\lambda L^2}}~.
\eeq
We notice that in the regime where 
$\eta^2 >  {{4 \pi^2 n^2}\over {\lambda L^2}}$ this is a
classically stable solution characterized by the topological number 
$n$, so one can consider the local minima labeled by $n$ 
as a set of ``flux vacua''. 

Even though these states are perturbatively stable they can decay by
quantum tunneling, which is described by an instanton that
interpolates between two states with different ``flux'' numbers. It is
clear that any such instanton would have to have at least one point in
the Euclidean spacetime where the phase of the complex scalar is
undefined. On the other hand, the Euclidean version of our original
Lagrangian (\ref{2D-action}) allows the possibility of vortex
solutions where the scalar field winds around the vortex center and
where the field $\phi$ goes to zero at the core.  It is therefore
reasonable to expect that the appropriate instanton would somehow
involve these solitonic solutions in Euclidean spacetime.

Indeed, the instanton solution that describes this kind of decay was
identified in \cite{String-tunneling} as a vortex and an anti-vortex
situated at different values of the Euclidean time (see Fig. 1). Using
our current notation, the instanton action is given by
\beq
\label{instanton-action}
B_E = 2 \pi \eta^2 \left(\ln\left({{2 d}\over \delta}\right) - {{2 \pi
    n}\over L} 2 d\right)
\eeq
where $\delta \sim {1\over {\sqrt{\lambda} \eta}}$ is the thickness of
the vortex core and $d$ is the distance between the two vortices in
Euclidean time, which is assumed to satisfy $d\ll L$.  The two terms
in Eq.~(\ref{instanton-action}) have simple physical interpretations.
The first term accounts for the self-energy of the vortices, with
the vortex separation $d$ providing the cutoff.  (We assume that the
logarithm in Eq.~(\ref{instanton-action}) is large, so the contribution
of the vortex core to the self-energy can be neglected.)
The second term takes into account the interaction of the
vortices with the background field (\ref{phiy}).\footnote{Notice that
in order to obtain the bounce action we have subtracted the
contribution from the original background flux.}  

One can see from this expression that the action is extremized when
\begin{figure}
\centering\leavevmode
\epsfysize=8cm \epsfbox{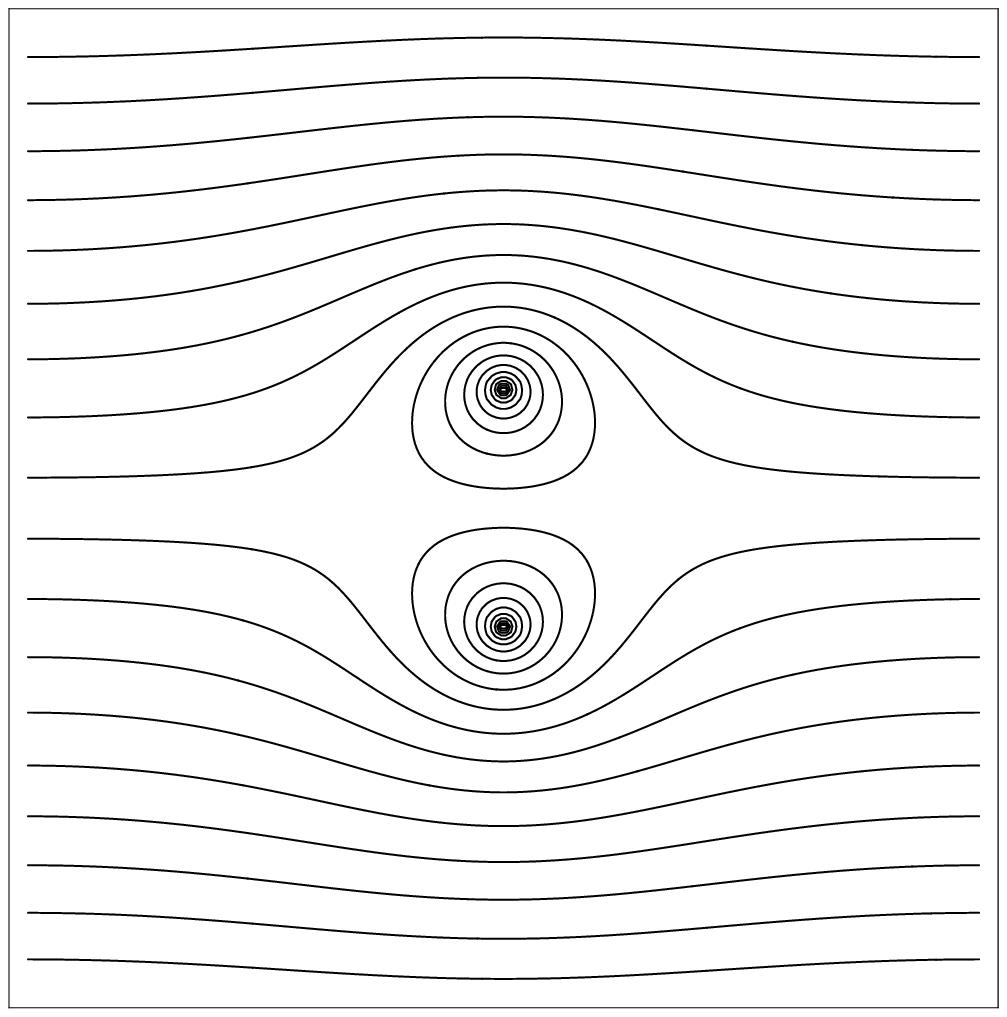}
\put(-210,-20){\Large {\bf {$y$}}}
\put(-190,-20){\Large {\bf {$\rightarrow$}}}
\put(-250,20){\Large {\bf {$\tau$}}}
\put(-250,40){\Large {\bf {$\uparrow$}}}
\put(10,110){\Large {\bf $\leftarrow$ $\tau=0$}}
\caption[Fig 1] {Instanton solution in the $1+1$ dimensional case. We
  plot the lines of constant $\chi$ defined in terms of 
the phase of the complex scalar field $\theta$  by the relation, 
{$\partial_a \chi = \epsilon_{ab}\partial_b \theta$}.}
\end{figure}

\beq
d = {L \over {4 \pi n}}~.
\eeq
This means that the distance between the vortices is always smaller than
the size of the $y$ dimension, $L$, and therefore we can use the
expression (\ref{instanton-action}) to calculate the instanton action

\beq
B_E = 2 \pi \eta^2 \left(\ln\left({{L}\over {2 \pi n \delta}}\right) - 1\right)~.
\eeq

The field configuration right after tunneling is given by the
instanton solution at $\tau = 0$.  It is clear from Fig.~1 that the
winding number of this configuration is one unit smaller than that
for the background solution.  The following evolution is obtained by
analytically continuing this configuration into the Lorentzian
regime.  It corresponds to a $1+1$ dimensional universe with one unit
of winding less than the initial state and two travelling pulses that
propagate the effects of the local instanton along the $y$-axis
(See \cite{String-tunneling} for details).

\subsection{$(2+1)$ dimensions}
We can now extend the previous discussion to the case where
we have one more spatial dimension with a Lagrangian of the form,

\beq
S_{2+1}=\int{dt dx dy \left(-  {1\over 2} \partial_{a} \phi
  {\partial^{a} {\bar \phi}} - {{\lambda}\over {4}} (\phi {\bar \phi} 
- \eta^2)^2\right)}~.
\label{3D-action}
\eeq

We can think of this model as a universe with one large spatial
dimension ($x$) and one compactified one ($y$). Similarly to the $1+1$
dimensional case, we can obtain static solutions for the scalar field
$\phi$ that wind around the extra dimension and that do not depend on
$x$. We can now imagine that we compactify the space along the $y$
direction, so the system effectively becomes $1+1$ dimensional.  This
is perhaps the simplest flux compactification one can think of, and
yet we shall see that it shares many of the relevant features with the
more realistic models that will be discussed later on.

We are interested in finding an instanton solution that allows the
configuration with a winding to decay. Such an
instanton should reduce to the configuration discussed in the previous
section when cut at a particular value of the large dimension $x$ but
clearly cannot be independent of $x$ since otherwise its action would
be infinite.  The solution is to consider the instanton made out of a
vortex string loop in Euclidean spacetime and located at a fixed value
of the extra dimension, $y$ (see Fig. 2). The slice at $\tau=0$ gives
the field configuration right after tunneling.  It describes a vortex
and an anti-vortex located at the same value of $y$ but separated by
some distance in the $x$ direction. After tunneling, the two vortices
move away from each other due to their interaction with the
background, which creates a Magnus force acting in opposite
directions on the two vortices. This leaves behind a growing region where
the winding number has been reduced by one unit\footnote{A similar
instanton has been found to be relevant in a very different context
for the case of quantum nucleation of strings loops in
\cite{KaoLee}.}

Having found the relevant instanton, we can compute, following a
similar argument as in the previous section, its Euclidean action. On
the other hand, it is worth looking at this model in its dual version,
where, as we will see, the calculation of the instanton Euclidean
action becomes completely straightforward. This is what we do next.
\begin{figure}
\centering\leavevmode
\epsfysize=9cm \epsfbox{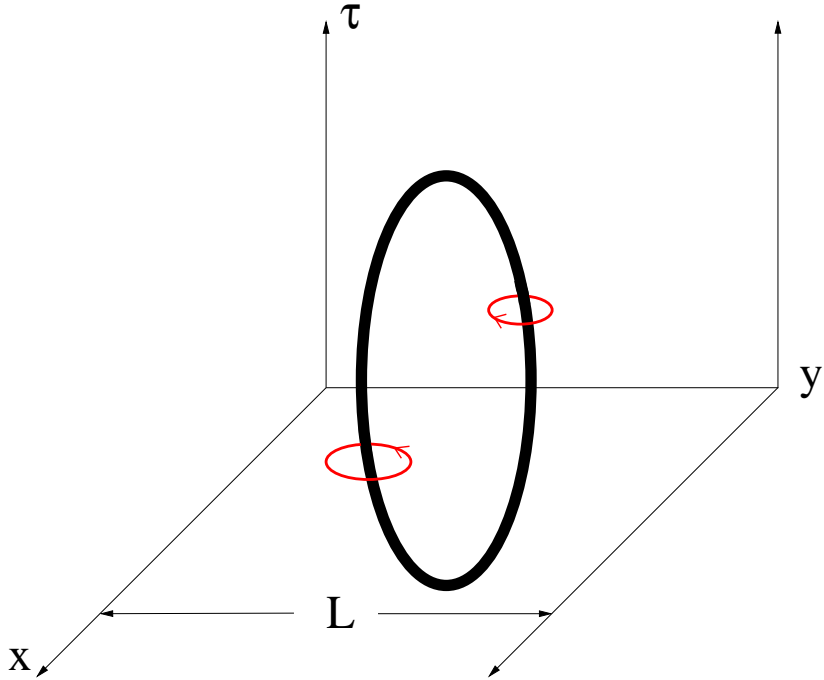}
\caption[Fig 2] {Vortex Ring Instanton. The arrows indicate the
  winding of the scalar field.}
\label{bubble-ring}
\end{figure}
\subsubsection{The dual $2+1$ dimensional theory}
We now want to describe the previous model in a slightly different
way by using a duality relation in $(2+1)$ dimensions between the 
phase of the complex scalar field $\theta$ and an electromagnetic 
field, $A_a$, namely,
\beq
F^{ab} = {{|\phi|^2}\over \eta} \epsilon^{abc} \partial_c \theta
\label{duality-2-1}
\eeq
where $a,b$ denote the 3 dimensional coordinates $t,x,y$; 
$F_{ab} = \partial_a A_b - \partial_b A_a$ and $\epsilon^{abc}$
is the totally antisymmetric tensor in $(2+1)d$. It is straightforward
to see that the equations of motion derived from the Lagrangian,

\beq
{\tilde S}_{2+1}=\int{dt dx dy \left(-  {1\over 2} (\partial_{a}
  |\phi|)^2- {{\lambda}\over {4}} (|\phi|^2 
- \eta^2)^2 - {\eta^2 \over {4|\phi|^2}} F^{ab} F_{ab}\right)}
\label{dual-3D-action}
\eeq
are the same as the ones obtained from (\ref{3D-action}), provided that
we use the duality relation (\ref{duality-2-1}). This also means that
the $2d$ vortices in the complex scalar field description should
now be identified with electrically charged point particles in the 
dual theory. 

At sufficiently low energies, when one freezes the scalar field $|\phi|
\sim \eta$, we can describe the effective theory in the dual picture
as\footnote{This is the same type of argument that was first
introduced for the effective description of global strings in a
$(3+1)$-dimensional theory in \cite{Davis-Shellard}.},
\beq
{\tilde S}_{2+1}=- m_v \int{ds \sqrt{-{{dx_a}\over{ds}} {{dx^a}\over {ds}}}}
+\int{dt dx dy \left(
- {1 \over {4}} F^{ab} F_{ab} + A_a J^a\right)}
\eeq
where $m_v$ represents the mass and  
\beq
J^a(x) = 2 \pi \eta \int{ds {{d x^a}\over {ds}} \delta^3(x - x(s))}
\eeq
is the 3-current associated with the charged particles in this
theory. We have identified the charge of the particle,
$q=2\pi\eta$, by making sure that we asymptotically get the same
solution for an isolated static vortex as in the scalar field theory.
This means that in the dual picture, and at low energies, our theory
is described by $(2+1)$ electromagnetism with pointlike
particles having a definite mass and charge, expressed in terms of the
parameters of the original theory.

Using Eq. (\ref{duality-2-1}) one can describe the original scalar
field winding in the dual picture as a constant electric field along the
uncompactified dimension $x$,
\beq
E_x =  \eta {{2 \pi n} \over {L}}~.
\eeq
We conclude from this that the tunneling process of winding decay in 
the scalar field theory can be thought of, via this duality,
as Schwinger pair production \cite{Schwinger} in $2+1$ dimensions. We can now
use this simple description to calculate
the Euclidean action for this process. Following \cite{Alex-nothing}
we note that the instanton can be thought of as a loop of a charged
particle worldline in Euclidean spacetime, so the action becomes
\beq
B_E = \pi \eta^2 \left(2 \pi R \ln\left({{2 R}\over \delta}\right) - \pi
R^2  {{4 \pi n}\over L} \right)~,
\eeq
where, as before, we have left the radius of the loop $R$ unspecified
and it should be found by extremizing the action. This happens roughly
when
\beq
R_E \sim {{L}\over {4 \pi n}} \ln \left({{L e}\over {2 \pi n
\delta}}\right),
\eeq
so the action turns out to be,
\beq
B_E \sim {{\pi \eta^2 L}\over {2 n}} \left(\ln \left({{L e}\over {2 \pi
    n \delta}}\right)\right)^2~.
\eeq

This estimate of the action disregards any effect due to
compactification, but it is clear that it has to be modified once the
radius of the loop is larger than the size of the extra dimension.  In
that limit, one would have to consider $L$ as the natural cutoff for
the logarithmic contribution of the effective mass of the vortex, so
the stationary value of the Euclidean action should be replaced by
\beq
B_E \sim {{\pi \eta^2 L}\over {2 n}} \left(\ln \left({{L }\over
{\delta}}\right)\right)^2~.
\eeq

In order to find the field configuration at nucleation, we note that
the charged particle loop in Euclidean spacetime can be seen as a
source for the Euclidean vector potential, much in the same way as a
loop of wire with a uniform current in Minkowski space\footnote{Note
that one can use the image method to obtain the correct boundary
conditions for $A^0$ that must be satisfied along the compact
dimension $y$.}. We have plotted in Fig. 3 the surfaces of constant 
$A^0(x,y)$, as well as the field lines for the configuration at $\tau=0$.

Having described the simplest scenario for tunneling in flux
compactification models, we now move on to spacetimes of higher
dimensionality that are much closer to realistic models of
compactification in string theory and other higher dimensional
theories. 
\begin{figure}
\centering\leavevmode
\epsfig{file=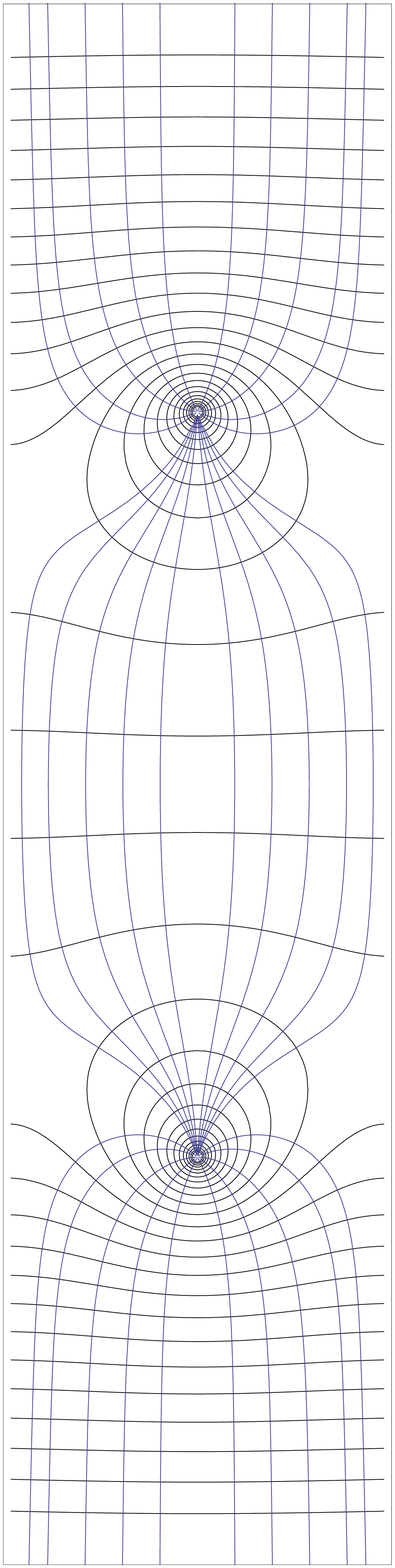,width=4cm,angle=90}
\put(-215,-20){\Large {\bf {$\rightarrow$}}}
\put(-480,75){\Large {\bf {$\uparrow$}}}
\put(-235,-20){\Large {\bf {$x$}}}
\put(-480,55){\Large {\bf {$y$}}}
\caption[Fig 3] {Vortex Ring Instanton at $\tau=0$. We plot the
  surfaces of constant potential as well as the field lines for the
pair production process in a constant electric field with  a compact
direction $y$.}
\end{figure}

\section{The Landscape of $5d$ Flux Compactifications}\label{The Landscape of 
Flux Compactifications}

In this section we would like to present a $5d$ scenario that shares
many of the properties of the toy models we discussed earlier, with the
important difference that we will now include gravity as a dynamical
part of our compactification. We consider an action of the form
\beq
S=\int{d^5\tilde x \sqrt{-\tilde g} \left(  {M_{(5)}^3\over 2}
  {\tilde R^{(5)}} -  {1\over 2} \partial_{M} \phi
  {\partial^{M} {\bar \phi}} - {{\lambda}\over {4}} (\phi {\bar \phi} 
- \eta^2)^2- {\tilde \Lambda}\right)},
\label{5D-complex-scalar-action}
\eeq
where $M,N,=0,...4$; $M_{(5)}$ denotes the $5d$ Planck mass, and we
have included a cosmological constant term ${\tilde \Lambda}$ which, as we will
see, is necessary to make the compactification in this type of model
possible. We note that, similarly to what we have seen already in
lower dimensional models, this kind of matter content allows the
possibility of codimension-2 solitonic objects, with the scalar field
winding around their core. These objects will be relevant to the
discussion of tunneling, but let us first describe the
compactification in these models.

\subsection{The $5d$ flux vacua}

For simplicity, we shall assume that the modulus of the scalar field
is effectively frozen at $|\phi|=\eta$, so that our effective action becomes,
\beq
S=\int{d^5\tilde x \sqrt{-\tilde g} \left(   {M_{(5)}^3\over 2}
  {\tilde R^{(5)}} - {1\over 2} \eta^2
  \partial_{M} \theta \partial^{M} \theta - {\tilde \Lambda}\right)},
\label{5D-action}
\eeq
where, similarly to the previous models, $\theta$ is the phase of $\phi(x^M)$.
The equations of motion for this model are:
\beq
\partial_M \left(\sqrt{-{\tilde g}} \partial^M \theta\right)=0~,
\label{scalar-eom}
\eeq
\beq
{\tilde R^{(5)}}_{AB} - {1\over 2} {\tilde g}_{AB} {\tilde R^{(5)}}
={1\over {M_{(5)}^{3}}} T_{AB}~,
\label{Einstein-eq-5D}
\eeq
where
\beq
T_{AB} = \eta^2 \left(\partial_A \theta \partial_B \theta - {1\over 2}
{\tilde g}_{AB} \partial_{M} \theta \partial^{M} \theta \right)
- {\tilde g}_{AB} {\tilde \Lambda}~
\eeq
is the total energy momentum tensor. We will look for a 
solution of the form
\beq
ds^2= {\tilde g}_{MN} dx^M dx^N = {\tilde g}_{\mu \nu} d x^{\mu}
d x^{\nu} + {\tilde g_{55}}(x^{\mu}) dx_5^2,
\label{5D-metric}
\eeq
where $\mu, \nu = 0,1,2,3$ denote the $4d$ coordinates and we assume
that the extra dimension has a compact range, 
$0 < x_5 < 2 \pi$. We are particularly interested in the 
case where
\beq
{\tilde g}_{55}(x^{\mu}) = L^2 = const~,
\eeq
in other words, in solutions where the extra dimension is stabilized
at a fixed radius $L$. We shall also require that the 4-dimensional
slices are described by a spacetime of constant scalar curvature
$R^{(4)} = 12 H^2$, where $H^2$ can be positive or negative, depending
on whether we are talking about de Sitter or anti-de Sitter
spacetime. With these assumptions, we arrive at the following five
dimensional Einstein tensor,
\begin{eqnarray}
{\tilde G}_{\mu \nu} &=& - 3 H^2 {\tilde g_{\mu \nu}}\\
{\tilde G}_{55} &=&  - 6  H^2 {\tilde g_{55}}~.
\end{eqnarray}

We are interested in solutions that resemble the flux
compactification examples we described before, so we impose
\beq
\theta(x^M) = n~x_5~.
\label{thetax5}
\eeq
The change of phase $\theta$ around the compact dimension should be an
integer multiple of $2\pi$; hence $n$ in Eq.~(\ref{thetax5}) should be
an integer.  Furthermore, given the form of our $5d$ metric, one can
see that this is in fact a solution of the equations of motion for the
scalar field, Eq. (\ref{scalar-eom}).

The energy momentum tensor becomes in this case,
\begin{eqnarray}
T_{\mu \nu} &=& - \left( {{n^2 \eta^2}\over{2 L^2}} + {\tilde \Lambda}\right)
{\tilde g}_{\mu \nu}, \\
T_{55} &=& \left({{n^2 \eta^2 }\over {2 L^2}} -  {\tilde
  \Lambda}\right){\tilde g}_{55}~.
\end{eqnarray}
Putting everything together, we arrive at the following equations for
$H$ and $L$:
\beq
3 H^2 = {1 \over {M_{(5)}^3}} \left( {{n^2 \eta^2}\over{2 L^2}} +
{\tilde \Lambda}\right),
\eeq
\beq
6 H^2  = - {1 \over {M_{(5)}^3}} \left({{n^2\eta^2}\over {2 L^2}} - {\tilde \Lambda}\right),
\eeq
which fix the values of $H$ and $L$ at:
\beq
L^2 = - {{3 n^2 \eta^2 }\over {2 {\tilde \Lambda}}},
\label{L2}
\eeq
\beq
H^2 = {2 {\tilde \Lambda} \over {9 {M_{(5)}^3}}}.
\label{H2in5D}
\eeq 

We conclude from Eq. (\ref{L2}) that it is possible to find a five-dimensional
solution with a compact extra dimension, provided that we start
with a $5d$ negative cosmological constant (${\tilde
  \Lambda} <0$). Eq.~(\ref{H2in5D}) then implies
that this compactification leads to a 4d anti-de Sitter spacetime.
On the other hand, the previous argument does not tell us anything 
about stability, in particular it is possible that the model is
already perturbatively unstable against small oscillations of the size
of the extra dimension. It is therefore useful to study this model
from a $4d$ point of view where one can identify the effective 
potential that controls the modulus that describes the size of the
compact space. We will do this in the next section.

\subsection{The 4d perspective}

It is possible to understand the origin of this compactification from
the dimensionally reduced effective theory in $4d$. Our starting point
is again the $5d$ action given by Eq. (\ref{5D-action}), namely, 
\beq
S=\int{d^5{\tilde x} \sqrt{-\tilde g} \left({{M_{(5)}^3} \over 2} 
  {\tilde R^{(5)}} - {1\over 2} \eta^2
  \partial_{M} \theta \partial^{M} \theta- {\tilde \Lambda}\right)}~.
\eeq
We can now take the $5d$ metric to be of the form,
\beq
ds^2= {\tilde g}_{MN} dx^M dx^N = e^{- \sqrt{{2\over 3}}
  \psi(x)/M_p} g_{\mu \nu} dx^{\mu} dx^{\nu} + e^{2\sqrt{{2\over 3}}
  \psi(x)/M_p} L^2 dx_5^2.
\label{5D-metric-2}
\eeq
Taking into account the solution for the scalar field
$\theta$ and the form of the metric, we can integrate the
5-dimensional action to get an effective theory in 4
dimensions written in terms of the field $\psi(x)$ as
\beq
S= \int{d^4 x \sqrt{-g}\left({1\over 2} M_p^2 R^{(4)} - {1\over 2}
  \partial_{\mu} \psi \partial^{\mu} \psi - V(\psi)\right)}.
\label{5d-dimensional-reduction}
\eeq
Where the $4d$ Planck mass is given by
\beq
M_p^2= 2 \pi L  M_{(5)}^3, 
\eeq
and the potential for the canonically normalized field $\psi$ is
given by

\beq
V(\psi, n) = 2 \pi L \left({\tilde \Lambda} e^{- \sqrt{{2\over 3}}
{\psi\over {M_p}}} +
\left({{\eta^2 n^2}\over{2 L^2}}\right) e^{- \sqrt{{6}} {\psi\over 
{M_p}}}\right). 
\eeq
\begin{figure}
\centering\leavevmode
\epsfysize=7cm \epsfbox{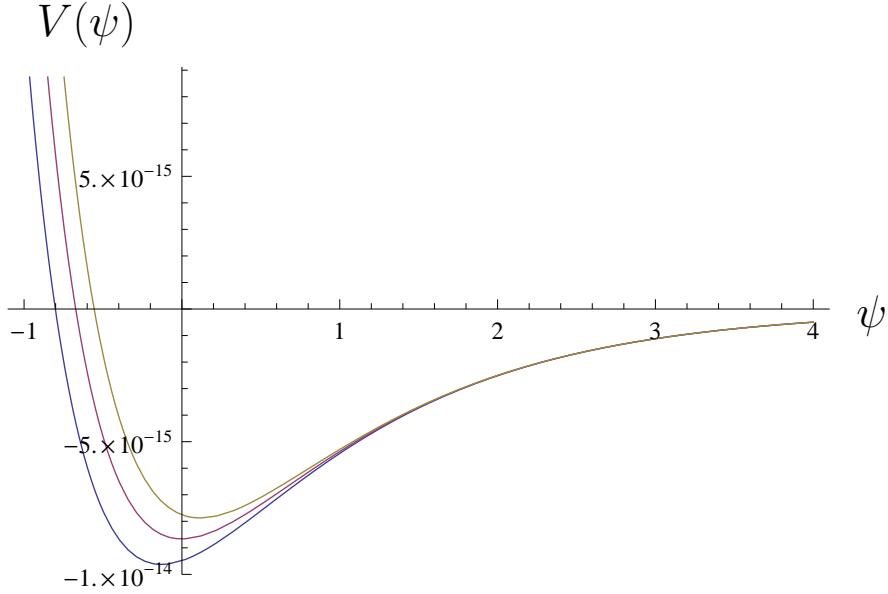}
\put(10,100){\Large {\bf {$\psi$}}}
\put(-300,210){\Large {\bf {$V(\psi)$}}}
\caption[Fig 1] {Plot of the $4d$ effective potential in $M_p$ units,
  as a function of the radius of the extra dimension for three
  different values of the winding number $n=9,10,11$. The
  parameters in the higher dimensional theory used here correspond to:
  $\eta^2 = 10^{-6} M_{(5)}^3$ and $\tilde \Lambda = -10^{-10}
  M_{(5)}^5$.}
\label{Potential}
\end{figure}
We see from this effective potential that it is only possible to
stabilize the field $\psi$ if one starts with a negative 5-dimensional
cosmological constant. For any particular vacuum $n_*$ we can always 
set the minimum of the potential at $\psi=0$ by identifying
\beq
L^2 = - {{3 \eta^2 n_*^2}\over {2 {\tilde \Lambda}}},
\eeq
so that the potential at the minimum becomes
\beq
V(\psi=0, n_*)= {{4 \pi L} \over 3} \tilde \Lambda = -4\pi \eta n_*
\sqrt{{- {{\tilde \Lambda}}\over {6}}}.
\eeq
We can now rewrite the potential for a general vacuum $n$,
using the previous definitions, as
\beq
V(\psi, n) = 2 \pi L \tilde \Lambda \left(e^{- \sqrt{{2\over 3}} {\psi\over {M_p}}} -
\left({{n^2}\over{3n_*^2}}\right) e^{- \sqrt{{6}} {\psi\over {M_p}}}\right)~.
\eeq
This potential is plotted in Fig.~4 for several values of $n$.

Finding the minima of this potential we can extract the spectrum of
cosmological constant values that $4d$ observers would be able to 
explore, namely
\beq
V(\psi_{min}, n) = {{4 \pi L} \over 3} \tilde \Lambda
\left({{n_*}\over {n}}\right)= -4\pi \eta n_* \sqrt{{- 
{{\tilde \Lambda}}\over {6}}}\left({{n_*}\over {n}}\right) 
\label{4d-cc}
\eeq
Notice that this is a special kind of landscape where all the 
values of the $4d$ cosmological constant that one is able to find
are negative. This is of course a limitation of the present toy model.

\subsection{The Dual version}

Similarly to what we did in the $(2+1)d$ case, we can also recast the
$5d$ model described above in terms of a four-form field, taking into
account that we now have the $5d$ duality relation,
\beq
{\tilde F}^{MNPQ} = {{\eta} \over \sqrt{-{\tilde g}}}~\epsilon^{MNPQR}~\partial_R \theta~.
\eeq
We can therefore rewrite the original action as
\beq
S=\int{d^5\tilde x \sqrt{-\tilde g} \left( {{M_{(5)}^3}\over 2}
{\tilde R^{(5)}} -{1\over {48}} {\tilde F}_{MNPQ} {\tilde F}^{MNPQ} - {\tilde
\Lambda}\right)}~.
\label{5D-action-dual}
\eeq
With this action, the equations of motion for the four-form are given by
\beq
\partial_M \left( \sqrt{-\tilde g} {\tilde F}^{MNPQ}\right) = 0~,
\label{fourform-equations}
\eeq
and Einstein's equations have the same form as we found before
in (\ref{Einstein-eq-5D}), except for the fact that we should use the
energy-momentum tensor for the four-form flux, namely,
\beq
T_{AB} = {1\over {4!}} \left( 4 {\tilde F}_{APQR} {{\tilde F}_B}^{PQR} - {1\over 2}
{\tilde g}_{AB} {\tilde F}^2 \right) - {\tilde g}_{AB} {\tilde \Lambda}~.
\eeq
Using the same ansatz for the metric that we had in
Eq.~(\ref{5D-metric}), the scalar field solution translates into
\beq
{\tilde F}^{\mu \nu \delta \gamma} = {1 \over{\sqrt{-{\tilde g}}}}
\epsilon^{\mu \nu \delta \gamma} n \eta , 
\label{5d-4-form-up}
\eeq
which also means that
\beq
{\tilde F}_{\mu \nu \delta \gamma} = \sqrt{-{\tilde g}}~
\epsilon_{\mu \nu \delta \gamma} \left({{n \eta }\over {L^2}}\right),
\label{5d-4-form-d}
\eeq
where $\mu, \nu, \delta, \gamma = 0,1,2,3$ and all the other
components are equal to zero. This is in fact 
a solution of Eq. (\ref{fourform-equations}) and leads to 
exactly the same energy momentum tensor as in the scalar field
description, so indeed we are just looking at the same $5d$ solution
in a somewhat different description.

One can see from (\ref{5d-4-form-up}-\ref{5d-4-form-d}) that
this solution corresponds to the excitation of only the zero mode of the 
$5d$ 4-form flux and therefore we should be able to understand this
landscape from a $4d$ theory of the form,

\beq
S= \int{d^4 x \sqrt{-g}\left({1\over 2} M_p^2 R^{(4)}-{1\over {48}}
  F_{\mu \nu \alpha \beta} F^{\mu \nu \alpha \beta} - \Lambda_{(4)}\right)}~
\label{4d-4-form-action}
\eeq
where $\sqrt{-\tilde g} = L \sqrt{-g}$, $M_P^2 = V_5 M^3_{(5)}$,
$\Lambda_{(4)} = V_5 {\tilde \Lambda}$ and $F_{\mu \nu \alpha \beta} =
\sqrt{V_{5}}{\tilde F}_{\mu \nu \alpha \beta}$ and $V_5 = 2 \pi L$. We
notice that this is the same type of action as the one studied a long
time ago in \cite{Brown-1,Brown-2}, where a {\it bare} negative
cosmological constant is compensated by the presence of the 4-form
flux contribution. With the relations specified above, we can
identify our $5d$ solution (\ref{5d-4-form-up}-\ref{5d-4-form-d}) in
the $4d$ language as

\begin{eqnarray}
 F^{\mu \nu \delta \gamma} &=& {1\over{\sqrt{-g}}}~
\epsilon^{\mu \nu \delta \gamma} \sqrt{{{2 \pi}\over {L}}} n \eta ~, \\
 F_{\mu \nu \delta \gamma} &=& \sqrt{-g}~
\epsilon_{\mu \nu \delta \gamma} \sqrt{{{2 \pi}\over {L}}} n \eta~.
\end{eqnarray}
We see that, similarly to what happens in the string theory case
\cite{Bousso:2000xa}, the $4d$ 4-form field strength is quantized. In
our example this requirement can be traced back to the dual
formulation of the theory where the quantization of the gradient of
the scalar field has a topological origin.

Finally, there seems to be a contradiction between the results for the
$4d$ cosmological constant using (\ref{4d-4-form-action}) and the ones
we obtained previously from the dimensionally reduced action for a
scalar field (\ref{5d-dimensional-reduction}). In particular, using
(\ref{4d-4-form-action}) one could conclude that it is possible to
balance the negative cosmological constant completely, so that the
$4d$ observers would be able to live in de Sitter or Minkowski
space. On the other hand, we have previously demonstrated, (see
(\ref{4d-cc})) that in this model all possible values of the
effective cosmological constant from the $4d$ perspective are in fact
negative. The reason for this discrepancy is that in
(\ref{4d-4-form-action}) we are disregarding the backreaction of the
fields on the geometry and assume a constant size of the extra
dimension for arbitrary values of $n$.  This deficiency in our $4d$
theory (\ref{4d-4-form-action}) can be remedied by incorporating the
size of the extra dimension as a degree of freedom of the low energy
theory, much in the same way as we did before. It is then clear that the
2-branes that are charged with respect to the 3-form potential in the
$4d$ language would also couple to this field, so that its value
would change across the domain wall in agreement with our higher
dimensional solutions.

\subsection{Another sector of the Landscape}

The dual formulation of our $5d$ model suggests that the same theory
could lead to a different sector of compactifications where the
spacetime is described by a $5d$ closed FRW type of universe. A 4-form
{\it monopole-like} flux can then be turned on the $S^4$ sphere of the
closed FRW manifold, allowing the possibility of different monopole
numbers as with the one-extra-dimensional case we just
studied\footnote{In its dual formulation this sector can be understood
as an example of the Freund-Rubin type of compactifications
\cite{Freund-Rubin}. On the other hand, in the original sector of the 
landscape, where we described the model in terms of scalars, one is
more inclined to think of the model as an example of spontaneous
compactification with scalars, like the ones discussed by Gell-Mann
and Zwiebach
\cite{GellMann}.}. One notices, however, that this type of flux
compactification requires a positive $5d$ cosmological constant in
order to have a static solution for the size of the internal manifold,
so there does not seem to be another sector of flux vacua in this model.

\subsection{Tunneling in the 5d model}

As we already mentioned, our model should really be thought of as the
low energy description of a complex scalar field living in a $(4+1)$d
universe. One can therefore expect the existence within this model of
solitonic solutions of codimension $2$ which are none other than
higher-dimensional generalizations of the vortex solutions described in
previous sections. It is then clear that one can use the same kind of
instanton solution to describe the flux tunneling in this case by
using these membrane\footnote{Note that these solutions of codimension
$2$ in a $(4+1)$-dimensional spacetime would have a
$(2+1)$-dimensional worldvolume, hence the name ``membrane''.}  solutions
instead of the string-like objects of Fig. 2.

In the dual description the branes also exist, although they are now
{\it electrically charged} objects with respect to a three-form 
potential. They do not appear in our action (\ref{5D-action-dual}), 
but, as in the lower dimensional cases described
above, one should supplement this action with the terms proportional to
their worldvolume as well as the coupling of the brane to the
four-form flux.

We have used the present model to visualize the instantons, but the
model clearly has an important limitation: the
$4d$ slices of spacetime are necessarily anti-de Sitter.
One can easily extend the ideas presented here to more
complicated models of higher dimensionality. This introduces new terms
in the low energy $4$d effective theory, which are proportional to the
curvature of the internal compactified manifold, so one can hope
to solve the problems present in the simplest scenario. Unfortunately,
we show in the Appendix that in fact the simplest generalizations of
this model to higher dimensions with scalar fields
compactified on a q-sphere are all unstable, unless the $4d$ universe
lives in anti-de Sitter space.

On the other hand, it is not difficult to find other models of
spontaneous compactification where one can circumvent this problem.
This is what we turn to in the following section.

\section{The Landscape of $6d$ Einstein-Maxwell theory}

\subsection{The flux vacua}

We will now discuss a $6d$ model, first proposed some time 
ago \cite{Freund-Rubin, EM6D}, that has recently received some attention
as a toy model for string theory compactifications
\cite{flux-compactifications}. The Lagrangian is
given by
\beq
S=\int{d^6 {\tilde x} \sqrt{-\tilde g} \left( {{M_{(6)}^4}\over 2}
{\tilde R}^{(6)} - {1\over 4} F_{MN} F^{MN} - {\tilde
\Lambda}\right)},
\label{EM-6D-action}
\eeq
where $M,N = 0...5$ label the six-dimensional coordinates, $M_{(6)}$
is the $6d$ Planck mass, and ${\tilde \Lambda}$ is the six-dimensional
cosmological constant.  The corresponding field equations are
\beq
{\tilde R}_{MN}^{(6)} - {1\over 2} {\tilde g}_{MN} {\tilde R}^{(6)} 
= {1\over {M_{(6)}^4}}  T_{MN}
\eeq
and
\beq
{1\over{\sqrt{-\tilde g}}} \partial_M \left(\sqrt{-\tilde g} F^{MN}\right)=0,
\eeq
with the energy-momentum tensor given by
\beq
T_{MN} ={\tilde g}^{LP} F_{ML} F_{NP} - {1\over 4} {\tilde g}_{MN} F^2 
- {\tilde g}_{MN} {\tilde \Lambda}.
\eeq 
We will look for solutions of this model with the spacetime metric
given by a four dimensional maximally symmetric space of constant
curvature,\footnote{As before, $H^2$ can be positive or negative,
depending on whether we are talking about de Sitter or anti-deSitter
spaces.} $R^{(4)} = 12 H^2$, and a static extra-dimensional 2-sphere
of fixed radius, namely a metric of the form,
\beq
ds^2= {\tilde g}_{MN} dx^M dx^N = {\tilde g}_{\mu \nu} d x^{\mu}
d x^{\nu} + R^2 d\Omega_2^2~.
\label{6D-metric}
\eeq 

With this ansatz, we obtain the following components of the $6d$
Einstein tensor,
\begin{eqnarray}
{\tilde G}_{\mu \nu}^{(6)} 
&=& -\left(3 H^2 + {1\over {R^2}}\right) \tilde g_{\mu \nu}\\
{\tilde G}_{ij}^{(6)} &=&  - 6 H^2 {\tilde g_{ij}} ~.
\end{eqnarray}
where we have used $\mu$ and $\nu$ to denote the 
four dimensional coordinates and $i$ and $j$ run over the
two extra dimensions on the sphere.

The only ansatz for the Maxwell field that is consistent with the
symmetries of the metric is a monopole-like configuration on the
extra-dimensional 2-sphere \cite{EM6D},
\beq
A_{\phi} = - {n \over {2 e}} (\cos \theta \pm 1) .
\eeq
Here, $n$ is an integer and the two signs denote the usual two
different patches necessary to describe the monopole field. The
quantization condition for $n$ comes from requiring that both
representations of the field must be related by a single-valued gauge
transformation along the equator of the sphere. The corresponding
field strength is easily computed to be
\beq
F_{\theta \phi} = -F_{\phi \theta} = {n \over{2 e}} \sin{\theta} ,
\eeq
which gives rise to the following energy-momentum tensor

\beq \label{Tmunuhdsol}
T_{\mu \nu} = - {\tilde g_{\mu \nu}} \left({{n^2}\over{8 e^2 R^4}} 
+ \tilde \Lambda\right)
\eeq
and
\beq \label{Tijhdsol}
T_{ij} = {\tilde g}_{ij} \left({{n^2}\over{8 e^2 R^4}} - \tilde
\Lambda\right).
\eeq
Putting everything together we arrive at

\beq
 3 H^2 + {1\over {R^2}} =  {1\over {M_{(6)}^4}}\left({{n^2}\over{8 e^2
     R^4}} + \tilde \Lambda\right)
\eeq
and

\beq
6 H^2 = {1\over {M_{(6)}^4}}\left(\tilde \Lambda - {{n^2}\over{8 e^2
R^4}}\right) .
\eeq

These equations are solved by

\beq
R^2 = {{M_{(6)}^4}\over{\tilde \Lambda}} \left(1 \mp \sqrt{1 - {{3 n^2
      \tilde \Lambda}\over {8 e^2 M_{(6)}^8}}}\right)
\eeq
and

\beq
H^2 = {{2 \tilde \Lambda}\over {9 M_{(6)}^4}} -  {{8 e^2 M_{(6)}^4}\over {27
    n^2}} \left(1 \pm \sqrt{1 -
  {{3 n^2 \tilde \Lambda}\over {8 e^2 M_{(6)}^8}}}\right) .
\eeq
We will see in the next section that only one of these solutions
is stable against small perturbations, so we will mostly be
interested in the upper signs in these equations.

\subsection{The 4d perspective}

It is interesting to understand the compactification mechanism from
the $4d$ perspective where the radius of the extra-dimensional
space becomes a dynamical field with a stabilizing potential.

Our starting point is again the higher dimensional theory,
Eq.~(\ref{EM-6D-action}). Following \cite{Garriga} we can now assume 
that the six dimensional metric has the form,
\beq
ds^2= {\tilde g}_{MN} dx^M dx^N = e^{- \psi(x)/M_P} g_{\mu \nu}
dx^{\mu} dx^{\nu} + e^{\psi(x)/M_P} R^{2}~d\Omega_2^2~.
\eeq
This ansatz, together with the monopole type configuration for the
Maxwell field, allows us to integrate the higher dimensional action 
over the internal manifold, to arrive at a $4d$ effective theory
of the form

\beq
S= \int{d^4 x \sqrt{-g}\left({1\over 2} M_P^2 R^{(4)} - {1\over 2}
  \partial_{\mu} \psi \partial^{\mu} \psi - V(\psi)\right)}.
\eeq
Here, the potential for the size of the internal dimension
is

\beq
V(\psi)= 4\pi M_{(6)}^4 \left({{n^2}\over{8 e^2 R^2 M_{(6)}^4}}
e^{-3\psi/M_P} - e^{-2\psi/M_P} + {{R^2 \tilde \Lambda} \over
{M_{(6)}^4}} e^{-\psi/M_P} \right)\label{4Deffectivepot} 
\eeq
and we have defined
\beq
M_P^2 = V_{S^2} M_{(6)}^4= 4 \pi R^2 M_{(6)}^4 \label{M_P} ,
\eeq
where $V_{S^2} = 4 \pi R^2$ is the area of a 2-sphere of radius $R$.

Once again, for any particular value of $n=n_*$ we can set
the minimum of the potential to be at $\psi =0$, by setting
\beq
R^2 = {{M_{(6)}^4}\over{\tilde \Lambda}} \left(1 - \sqrt{1 - {{3 n_*^2
\tilde \Lambda}\over {8 e^2 M_{(6)}^8}}}\right).
\label{R2}
\eeq
The value of the potential at this minimum is then given by
\beq
V(\psi=0,n_*)={{4\pi M_{(6)}^4}\over {3}} \left(1- 2 \sqrt{1 - {{3 n_*^2
\tilde \Lambda}\over {8 e^2 M_{(6)}^8}}}\right)~.
\eeq
We can now use the definition of the Planck mass in four dimensions to
calculate the Hubble expansion rate that a four-dimensional observer
would see while sitting at the minimum of the potential, namely,

\beq
H^2 = {{V(\psi=0,n_*)}\over {3 M_P^2}} = {{2 \tilde \Lambda}\over
 {9 M_{(6)}^4}} -  {{8 e^2 M_{(6)}^4}\over {27 n_*^2}} \left(1 + \sqrt{1 -
  {{3 n_*^2 \tilde \Lambda}\over {8 e^2 M_{(6)}^8 }}}\right),
\eeq
which is, of course, the same expression as we found from the
higher-dimensional theory.

One can fix the value of the potential at the minimum to be zero at
some $n=n_0$ by imposing the following relation 
\beq
{{n_0^2 \tilde \Lambda}\over {e^2 M_{(6)}^8 }} = 2~.
\label{n0}
\eeq
The radius of the compact dimensions in the corresponding vacuum
is given by
\beq
R_0={{M_{(6)}^2}\over{\sqrt{2 {\tilde\Lambda}}}}~.
\label{R0}
\eeq

\begin{figure}
\centering\leavevmode
\epsfysize=8cm \epsfbox{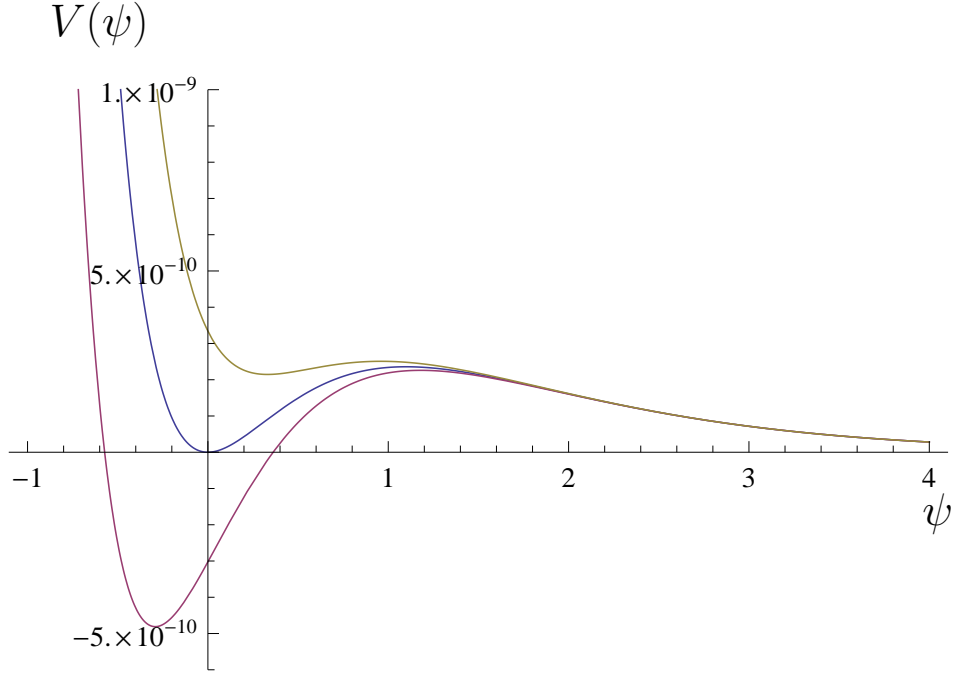}
\put(-10,55){\Large {\bf {$\psi$}}}
\put(-340,240){\Large {\bf {$V(\psi)$}}}
\caption {Plot of the $4d$ effective potential, in $M_P$ units,
as a function of the field $\psi$. We show the potential for 3
different values of the flux quantum $n = 180, 200, 220$. The rest of
the parameters of the model are fixed according to the relations given
in the main text.  }
\label{6D-potential}
\end{figure}

For other values of $n$, the effective potential has the form
\beq \label{potential}
V(\psi,n, n_0)= 4\pi M_{(6)}^4 \left({{n^2}\over{2 n_0^2}}
e^{-3\psi/M_p} - e^{-2\psi/M_P} + {1 \over 2} e^{-\psi/M_P} \right) .
\eeq

The minima of this potential for different values of $n$ will
constitute a ``landscape'' of vacua with different values for the
effective cosmological constant in the $4d$ theory, given by

\beq
V(\psi_{min},n,n_0)=4 \pi M_{(6)}^4 {\gamma \over 4} \left[1- {2\over
3} \gamma \left( 1 + \left(1 - {1\over
\gamma}\right)^{3/2}\right)\right] ,
\eeq
where we have defined $\gamma = {{4 n_0^2}\over {3 n^2}}$.
The boundaries of this toy landscape are determined by the positivity
of the expression under the square root in
Eq.~(\ref{R2}):\footnote{Note that this agrees with the stability
analysis of \cite{Bousso-deWolfe}.}
\beq
n^2\leq {4\over{3}}n_0^2.
\eeq
Hence, in order to have a large landscape, we need $n_0\gg 1$, or
\beq
{\tilde\Lambda}\ll e^2 M_{(6)}^{8}.
\eeq

\begin{figure}
\centering\leavevmode
\epsfysize=7cm \epsfbox{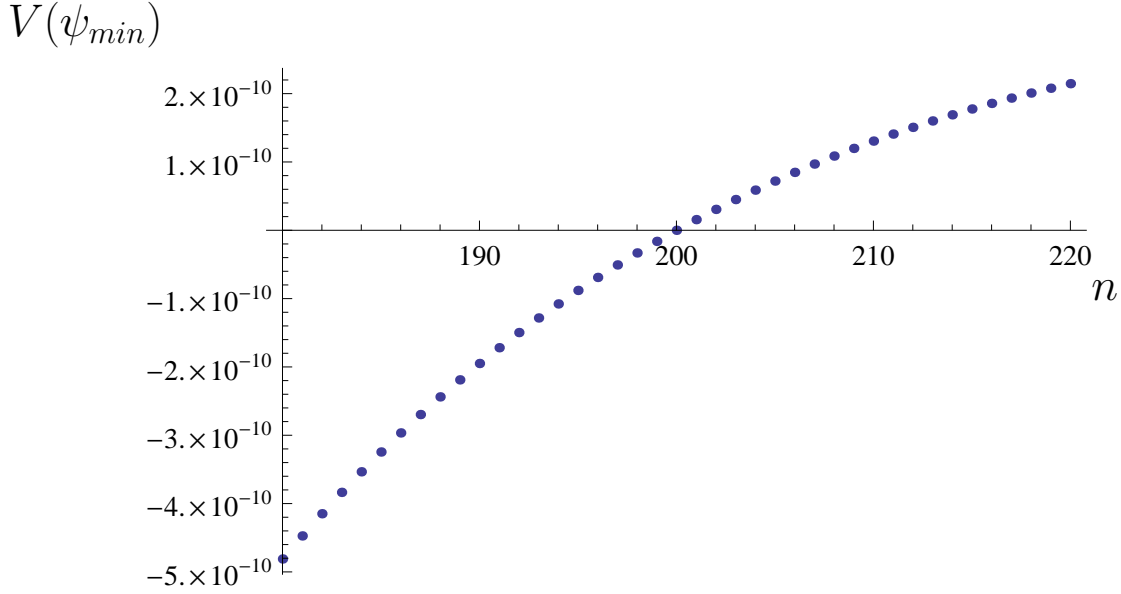}
\put(0,110){\Large {\bf {$n$}}}
\put(-410,210){\Large {\bf {$V(\psi_{min})$}}}
\caption{Values of the cosmological constant in units of
$M_P^4$ for $180 < n < 220$.}
\label{6D-landscape}
\end{figure}

As an illustrative example, we consider the values $e^2 M_{(6)}^2 = 2
$ and ${\tilde \Lambda}/M_{(6)}^6 = 10^{-4}$.  The condition of
vanishing cosmological constant is then satisfied for $n_0 = 200$. We
plot in Fig. \ref{6D-potential} the effective potential for $n
=180,200,220$. We also plot in Fig. \ref{6D-landscape} the values of
the cosmological constant in the range from $n=180$ to $n=220$. The
jumps in energy density between the adjacent vacua in this range are
nearly constant and are given by

\beq
\Delta V \approx {\partial V\over{\partial n}}(n_0,\psi=0) 
\approx {{4 \pi M_{(6)}^4}\over n_0} .
\label{DeltaV}
\eeq

\section{Tunneling in the Einstein-Maxwell theory}

It is clear from Fig. \ref{6D-potential} that for any given value of
$n$, within the range shown there, one has stable vacua under small
perturbations in the compactification radius. We also notice that the
potential tends to zero for large values of the radius, which in turn
means that positive-energy vacua should be able to decay by tunneling
through a barrier, leading effectively to decompactification of
space. This seems to be a generic situation for the four dimensional
effective potentials for the moduli that represent the size of the
internal manifold and that are stabilized at a non-negative value of
the $4d$ cosmological constant \cite{KKLT,Decompactification}. It is
interesting to estimate the decay rate of the above vacua towards
decompactification and compare it with other means of decay.

We note that decompactification can also occur via quantum
diffusion. This was studied in a $6d$ related model in
\cite{LindeZelnikov}. In the present paper we will concentrate
on quantum tunneling events.

\subsection{Decompactification tunneling}

Decompactification tunneling can be described using either the
Hawking-Moss \cite{HM} (HM) instanton or the Coleman-De Luccia
\cite{CdL} (CDL) instanton, depending on the form of our potential.

The CDL formalism applies if
\beq 
\label{domainwallbound}
|V'' (\psi_{max})|^{1/2} > 2 H_{max} ~,
\eeq
where $\psi_{max}$ is the value of $\psi$ at the maximum of $V(\psi)$
in the potential barrier and $H_{max}$ is the corresponding value of
$H$.  In this regime, the vacua inside and outside the bubble are
separated by domain walls of fixed thickness \cite{Rama}.
Alternatively, if (\ref{domainwallbound}) is not satisfied, the domain
walls are inflating \cite{topinf1,topinf2}, and we are in the Hawking-Moss
regime.

Let us first consider $n \approx n_0$, where $n_0$ is
the value for which the vacuum energy vanishes, given by
Eq.~(\ref{n0}).  Using Eq.~(\ref{potential}) for $V(\psi)$, we find
that in this case $\psi_{max} \approx M_P \ln 3$, $V(\psi_{max})
\approx {8\pi M_{(6)}^4 } / 27$,
\beq
H_{max} \approx {2  M_{(6)}^2\sqrt{2 \pi} \over 9M_P} ,
\eeq
and
\beq
|V'' (\psi_{max})|^{1/2} \approx {{2 \sqrt{\pi} M_{(6)}^2}\over {3
M_P}}.
\eeq
Thus, we find
\beq
|V'' (\psi_{max})|^{1/2} \approx {3 \over \sqrt2} H_{max}>2 H_{max}.
\eeq

Notice that Eq.~(\ref{domainwallbound}) is marginally satisfied. As
$n$ increases, the potential barrier becomes flatter, and we
expect to shift away from this marginally CDL regime, into the HM
regime. We expect, therefore, that for all values of $n$ the
tunneling action is well approximated by that for the Hawking-Moss
instanton.

The HM action is given by
\beq
S_{HM}=24\pi^2 M_P^{4} \left({1\over{V_{min}}}-{1\over{V_{max}}}\right)
\eeq
where $V_{min}$ is the vacuum energy density prior to the tunneling.
For $n = n_0 + m$ with $m\ll n_0$, we have $V_{min}=m \Delta V$, where
$\Delta V$ is given by Eq.~(\ref{DeltaV}).  Using Eqs. (\ref{M_P}),
(\ref{R0}) and (\ref{n0}), we can rewrite the action as

\beq
S_{HM} \approx {24\pi^2 M_P^{4} \over {\Delta V}}{1 \over m}
\left(1-{27 m \over 2 n_0}\right) \\
\approx {{6 \pi^3 n_0^5}\over{(eM_{(6)})^4 m}}
\label{SHMcompare}
\eeq
For the model parameters used in section IV,
we have $S_{HM} \approx 1.4 \times 10^{13}$.

\subsection{Flux tunneling}

We shall now argue that the instanton that interpolates between
compactifications with different flux quantum number $n$ has the form of
a {\it bubble ring} in the $6d$ Euclidean spacetime. One can think of
this object as an $O(4)$ symmetric bubble living at some fixed
values of the extra-dimensional coordinates on the sphere.  This is a
codimension 3 object, one of these dimensions is along the radial
direction $\rho=
\sqrt{\tau^2+ x_1^2 +x_2^2+ x_3^2}$ and the other two are the internal
directions on the 2-sphere. Since the tunneling has to reduce the
magnetic flux, this object has to be magnetically charged with respect
to our Maxwell field, and symmetry dictates that the magnetic
flux crossing any 2-sphere that surrounds the object should be
constant. In other words, this object should be an extended version of
a magnetic monopole in $4d$. We can borrow the string theory language
here and name this object a $2$-brane, a membrane of 2 intrinsic
dimensions magnetically charged with respect to the potential
$A_{M}$. The physical origin of these branes in our model will be
discussed in the following subsection.

Taking into account the properties of this 2-brane described above we
can convince ourselves that the {\it bubble ring} instanton is indeed
the correct Euclidean solution we are looking for. In fact, if we look
at the solution at $\tau = 0$ we see that this is precisely what we
need, since inside of the bubble the magnetic flux through extra
dimensions has now been reduced by the unit magnetic charge of the
brane.

In the vicinity of the brane, at distances much smaller than the
compactification radius $R$, the magnetic field of the brane is nearly
spherically symmetric, like the field of a magnetic monopole.  On the
other hand, at large distances from the brane the field should
approach a vacuum solution.  If the bubble radius is
\beq
\rho\gg R,
\label{thinwall}
\eeq
then the $4d$ regions inside and outside the bubble are nearly
homogeneous vacua, differing by one unit of magnetic flux through the
compactified dimensions.  The transition between the two regions
occurs in a shell of thickness $\Delta\rho \sim R$, which plays the
role of the domain wall.  The condition (\ref{thinwall}) corresponds
to the thin wall regime.

In the opposite regime, the initial size of the bubble is small
compared to the size of extra dimensions, $\rho\ll R$, and the
nucleation probability can approximately be found by considering the
limit $R\to\infty$.  The instanton then describes the nucleation of a
spherical 2-brane in a constant external field in $6d$.  This is a
higher-dimensional analogue of the nucleation of monopole-antimonopole
pairs in a homogeneous magnetic field \cite{monopole-pair-production}.

\subsubsection{2-brane solutions}

Our original Lagrangian did not have any branes of the form that
we conjectured in the instanton, but we will now show that one can find
solutions that describe these types of objects in the $6d$
Einstein-Maxwell theory.

We are interested in an object that has
Lorentz symmetry along $2+1$ dimensions and is spherically
symmetric with respect to the perpendicular directions. In
this case we can write the most general solution for gravity as

\beq
ds^2 = A(r)^2 (-dt^2 + dx^2 + dy^2) + B(r)^2 dr^2 + r^2 d\Omega_2^2 .
\eeq
We want this object to be magnetically charged, so we take the 
solution for the electromagnetic field strength to be,
\beq
F_{\theta \phi} = {g \over {4 \pi}} \sin{\theta}
\eeq
which gives rise to the following energy-momentum tensor:

\beq
T^{\mu}_{\nu} = - {1\over 2}\left({{g}\over {4 \pi 
    r^2}}\right)^2 \delta^{\mu}_{\nu} 
\eeq
and
\beq
T^{i}_{j} = {1\over 2}\left({{g}\over {4 \pi 
    r^2}}\right)^2  \delta^i_j~.
\eeq
where $\mu, \nu = t,x,y,r$ and $i,j = \theta, \phi$.  Similarly to
what happens for the four-dimensional Reissner-Nordstrom black holes,
one finds that there is a two-parameter family of solutions for
magnetically charged branes. These type of solutions can be found 
in \cite{Gibbons} and can also be obtained by
taking the appropriate limit that decouples the dilaton field in the
solutions found in \cite{Gregory}.

In the following we concentrate on the extremal case where one can
write the solution in terms of a single parameter, $r_0$. The solution
in this limit becomes

\beq
A(r)^2 = \left(1- {r_0\over r}\right)^{2/3} ,
\eeq
\beq
B(r)^2 =  \left(1- {r_0\over r}\right)^{-2} 
\eeq
and
with
\beq
r_0 =  {{\sqrt{3}~g} \over {8 \pi M_{(6)}^2}} .
\eeq
Thus we finally arrive at the solution of the form

\beq
ds^2 = \left(1- {{\sqrt{3}~g}\over {8 \pi M_{(6)}^2 r}}\right)^{2/3} (-dt^2
+ dx^2 + dy^2) + \left(1- {{\sqrt{3}~g}\over {8 \pi M_{(6)}^2 r}}
\right)^{-2} dr^2 + r^2 d\Omega_2^2.
\eeq
We are interested in the minimally charged brane which will
interpolate between consecutive flux vacua. We can see from the 
definition of the field strength on the 2-sphere compactification
in the previous section that this imposes

\beq
g e = 2 \pi
\eeq
which is, of course, the generalization of Dirac's condition to our
six dimensional model.

We can now compute the tension of these branes from the asymptotic form
of the metric. Following the calculations in \cite{Lu} we obtain,

\beq
T_{2} = {{16 \pi}\over {3}} M_{(6)}^4 r_0 = {{2g M_{(6)}^2}\over {\sqrt{3}}}
\label{T2}
\eeq

It is important to notice that the energy-momentum tensor associated
with the magnetic charge is fairly localized, decaying quite fast with
the radial distance from the brane. In fact, most of the energy of
these branes is concentrated in a region of the order $r_0$ around the
brane core.

The solution presented here is just one particular limit, the extremal
case, of a family of solutions with the same magnetic charge but
different tension \cite{Gregory}.  One may then wonder what tension one should use to
compute the instanton action in our flux tunneling decay. One way to
resolve this issue is to embed the abelian monopole compactification
we have been discussing in this section in a non-abelian
Einstein-Yang-Mills-Higgs model like the one studied in
\cite{Cremmer}. That type of model would have smooth solitonic
magnetically charged solutions (2 branes) that could be used to
construct the tunneling instantons that we are interested in.  The
tension of these branes will be fixed in terms of the underlying
parameters in the field theory, therefore selecting one particular 
element of the 2-parameter family described above.

\subsubsection{The instanton action}

The instanton action can be easily found in the thin wall limit, when
the radius of the bubble ring is much greater than the 
compactification radius,
\beq
\rho\gg R.
\label{rho>>R}
\eeq  
We shall use the standard Coleman-De Luccia formalism \cite{CdL,Parke}
to consider tunneling between the following types of vacua:  de Sitter
to de Sitter;  de Sitter to Minkowski; and from a Minkowski vacuum to 
the nearest AdS vacuum.

For the general case of tunneling between vacua, the bubble radius
$\rho$ and the tunneling action $S_E$ can be
expressed as \cite{Parke}
\beq
\rho=\rho^{(0)}[1+2xy+x^2]^{-1/2},
\label{rhoCdL}
\eeq
\beq
S_E=S_E^{(0)}r(x,y).
\label{SECdL}
\eeq
Here, $\rho^{(0)}$ and $S_E^{(0)}$ are the corresponding flat-space
expressions \cite{Coleman}, obtained neglecting the effects of $4d$
gravity and are given by:

\beq
\rho^{(0)}= 3\sigma/\Delta V,
\label{rho}
\eeq
\beq
S_E^{(0)} = {27\pi^2 \sigma^4\over{2(\Delta V)^3}},
\label{Sthinwall}
\eeq
$\sigma$ is the domain wall tension, and $\Delta V$ is the energy
density difference between the two vacua (see Eq. (\ref{DeltaV})).  We
have also defined

\beq
x={{3 \sigma^2}\over{4 M_P^2 \Delta V}},
\label{x}
\eeq

\beq
y={2 V_{initial} \over \Delta V} -1,
\label{y}
\eeq

and the gravitational factor is
\beq
r(x,y)={{2[(1+xy)-(1+2xy+x^2)^{1/2}]}\over{x^2(y^2-1)(1+2xy+x^2)^{1/2}}} .     
\eeq 

For the special cases of tunneling from de Sitter to Minkowski ($y=1$)
and Minkowski to AdS ($y=-1$),

\beq
\rho=\rho^{(0)}[1\pm x]^{-1},
\label{rhoCdLspecial}
\eeq

\beq
S_E=S_E^{(0)}[1\pm x]^{-2}.
\label{SECdLspecial}
\eeq
where the plus sign is for dS-Minkowski, and the minus sign is for
Minkowski-AdS.

Note that the curvature scale of the AdS vacuum inside the bubble is
\beq
|H|=(|V_{AdS}|/3M_P^2)^{1/2}
\eeq 
with
\beq
|V_{AdS}|=\Delta V \approx 4\pi M_{(6)}^4/n_0.
\eeq
Also, as we discussed, in the thin wall limit the energy of the wall
is concentrated mainly in the brane.  Hence, we can write
\beq
\sigma\approx T_2. 
\label{sigma}
\eeq
and, using Eqs.~(\ref{M_P}), (\ref{n0}) and (\ref{R0}), 
\beq
x={3\over{4 n_0}}\left({T_2\over{M_{(6)}^2 g}}\right)^2.
\eeq
For extremal branes with $T_2$ given by (\ref{T2}) and a large
landscape with $n_0\gg 1$, this gives
\beq
x = {1\over { n_0}} \ll 1.
\eeq
In this case, the gravitational corrections in
(\ref{rhoCdL}), (\ref{SECdL}), (\ref{rhoCdLspecial}) and
(\ref{SECdLspecial}) are negligible, and we can use the flat
space relations (\ref{rho}), (\ref{Sthinwall}).  
To test the validity of the thin wall condition (\ref{rho>>R}), 
we consider the ratio
\beq
{\rho\over{R_0}} \approx {3T_2 n_0\sqrt{2{\tilde\Lambda}}\over{4\pi
    M_{(6)}^6}} = {3T_2\over{M_{(6)}^2 g}},
\eeq
where in the last step we used the relation (\ref{n0}).  For extremal
branes, $\rho/R_0 = 2\sqrt{3}$, and the condition (\ref{rho>>R}) is
only marginally satisfied, but one can expect that the thin-wall
expression for the action (\ref{Sthinwall}) is still valid by
order of magnitude.  Then,
\beq
S_E \sim S_E^{(0)}= {{24\pi^2 x^2 M_P^4}\over{\Delta V}} 
\sim {3\over {8 \pi}} \left({g\over {M_{(6)}}}\right)^4 n_0^3.
\label{Sfluxextr}
\eeq

By comparing Eq.~(\ref{Sfluxextr}) with Eq.~(\ref{SHMcompare}), one
can immediately see that vacuum decay via decompactification is strongly
suppressed compared to that via flux tunneling.

For superheavy branes with $T_2\gtrsim M_{(6)}^2 g\sqrt{n_0}$ the
effects of gravity become important; they completely suppress vacuum
decay from Minkowski to AdS vacua for $x
\geq 1$. For $x>>1$, we find

\beq
r(x,y) \approx {2 \over {x^2 (1+y)}}
\eeq
and
\beq
S_{superheavy}\approx {{24\pi^2 M_P^4}\over{V_{min}}} , 
\label{Ssuperheavy}
\eeq
where $V_{min}$ is the potential energy density in the initial vacuum from
which we are tunneling.  Notice that in this large tension regime, the
tunneling action is independent of the tension.  Note also that 
Eq.~(\ref{Ssuperheavy}) is approximately the same as the
decompactification tunneling action, Eq.~(\ref{SHMcompare}), so the
decay rates into these two channels should be comparable.
 
Apart from the thin wall regime, the tunneling action can also be
estimated in the opposite limit, when $\rho\ll R$.  This is more
conveniently done in the dual picture, to which we shall now turn.

\section{The Dual Picture}

Once again we can recast the $6d$ model described above in terms of a four-form
field, using the duality relation,

\beq
\tilde F^{MNPQ} = {1\over {2 \sqrt{-{\tilde g}}}}~{\epsilon^{MNPQRS}}~F_{RS}~.
\eeq
The action for this model becomes

\beq S=\int{d^6\tilde x \sqrt{-\tilde g} \left({{M^4_{(6)}}\over 2}
  {\tilde R} -{1\over {48}} \tilde F_{MNPQ} \tilde F^{MNPQ} - {\tilde
    \Lambda}\right)}~.
\label{6D-dual-action}
\eeq
The corresponding  equations of motion are

\beq
\partial_M \left( \sqrt{-\tilde g} \tilde F^{MNPQ}\right) = 0~,
\label{4-form-eom}
\eeq

\beq {\tilde R}_{AB} - {1\over 2} {\tilde g}_{AB} {\tilde R} = {1\over
  {M^4_{(6)}}} T_{AB}~, \eeq 
and the energy momentum tensor is given by

\beq T_{AB} = {1\over {4!}} \left( 4 \tilde F_{APQR} {\tilde
  F_B}^{PQR} - {1\over 2} {\tilde g}_{AB} \tilde F^2 \right) - {\tilde
  g}_{AB} {\tilde \Lambda}.  
\eeq 
Using the same ansatz for the metric
as before, namely Eq. (\ref{6D-metric}), the monopole-like
configuration becomes

\beq
\tilde F^{\mu \nu \delta \gamma} =  {{\epsilon^{\mu \nu \delta
      \gamma}}\over{\sqrt{-{\tilde g}}}}
~\left({{n}\over {2 e}} \sin \theta \right) =  {{\epsilon^{\mu \nu \delta
      \gamma}}\over{\sqrt{-{\tilde g_4}}}}
~\left({{n}\over {2 e R^2}} \right)
\label{Fdual}
\eeq
where $\mu, \nu, \delta, \gamma$ denote only the $4d$ indices, $\tilde
g_4$ is the determinant of the $4d$ part of the higher dimensional
metric (\ref{6D-metric}), and all the other components of the 4-form
tensor are equal to zero. This is in fact 
a solution of Eq. (\ref{4-form-eom}) and leads to 
exactly the same energy momentum tensor, Eq.'s  (\ref{Tmunuhdsol}) 
and (\ref{Tijhdsol}), as before.

The action (\ref{6D-dual-action}) should be supplemented by the brane
action, 
\beq
S_{brane} = -T_2 \int d^{(3)}\Sigma + {g\over{3!}} \int \tilde A_{MNP}d^{(3)}
\Sigma^{MNP},
\label{Sbrane}
\eeq
where the second term describes the coupling of the brane to the form
field, $g$ is the corresponding charge\footnote{Recall
  that the {\it electrically} charged branes in the 4-form formalism
in $6d$  correspond to the magnetically charged branes in the model described
  in terms of the Maxwell field.  This is why we use $g$ to denote the
  charge of the branes in this version of the model.}, and the
potential $\tilde A_{MNP}$ is related in the usual way to the field strength by
\beq
\tilde F_{MNPQ}=\partial_{[M}\tilde A_{NPQ]}.
\eeq
The integration in (\ref{Sbrane}) is over the 3-dimensional
worldsheet of the brane.

\subsection{The instanton in the dual description}

The structure of the instanton in the dual picture is essentially
unchanged, with the replacement of the field $F_{MN}$ by its dual
four-form field, which is now {\it electrically} coupled to the brane.  The
instanton action in the thin wall limit can be analyzed along the same
lines as before, so we shall not discuss it here.  Instead, we shall
consider the opposite limit of a small bubble ring,
\beq
\rho\ll R.
\label{rho<<R}
\eeq
As we already mentioned, this regime can be studied by letting
$R\to\infty$.  The instanton then describes nucleation of spherical
branes in a constant external field (\ref{Fdual}).

We shall estimate the action of this instanton in the test brane
approximation, that is, assuming that the brane has only a small
effect on the background geometry and the four-form field.  Here
we shall assume that the initial vacuum has zero cosmological constant.
The action can then be found from the brane action (\ref{Sbrane}) in
flat space and treating $\tilde A_{MNP}$ as an external field.  The
contribution of the first term in (\ref{Sbrane}) is $T_2 \Sigma_3$,
where $\Sigma_3 = 2 \pi^2 \rho^3$ is the volume of a 3-sphere (which
is the Euclidean worldsheet of the brane).  

The second term can be evaluated using the Stokes theorem,
\beq
\int_\Sigma \tilde A_{MNP}d^{(3)}\Sigma^{MNP} = {1\over{4}}
\int_\Omega \tilde F_{MNPQ}d^{(4)}\Sigma^{MNPQ},
\eeq
where $\Omega$ is a 4-dimensional surface which is bounded by the
3-dimensional surface $\Sigma$.  Taking $\Sigma$ to be our spherical
bubble worldsheet, we obtain
\beq
{g\over{3!}}\int \tilde A_{MNP}d^{(3)}\Sigma^{MNP} = g \tilde F\Omega_4 ,
\eeq
where $\Omega_4 = (\pi^2/2)\rho^4$ is the 4-volume enclosed by the
3-sphere and ${\tilde F}$ is the field strength (the factor multiplying 
$\epsilon_{\mu\nu\sigma\tau}$ in Eq.~(\ref{Fdual})).  Assuming
that the initial vacuum is close to Minkowski, $n\approx n_0$, we have 
\beq
\tilde F \approx {{n_0}\over {2eR_0^2}}=\sqrt{2{\tilde\Lambda}} 
\label{F}
\eeq

Combining the two terms, we obtain
\beq
S_E = T_2 \Sigma_3 - g \tilde F\Omega_4 = 2 \pi^2 T_2 \rho^3 - {\pi^2\over{2}}
g\tilde F\rho^4.
\label{SEbrane}
\eeq
The bubble radius can now be found by minimizing this with respect to
$\rho$,
\beq
\rho ={3T_2\over{g \tilde F}}.
\eeq
Substituting this back to the action, we get
\beq
S_E ={27\pi^2\over{2}} \left({T_2^4\over{g^3 \tilde F^3}}\right).
\label{Stestbrane}
\eeq

To check the validity of the small bubble condition (\ref{rho<<R}), we
evaluate
\beq
{\rho\over{R_0}} = {3 T_2\over{g M_{(6)}^2}}.
\eeq
As before, the extremal brane tension (\ref{T2}) corresponds to the
marginal case, $\rho\sim R_0$, while the small bubble condition
requires that $T_2\ll g M_{(6)}^2$.

The test brane approximation is justified if the force on the brane
due to the external field $\tilde F$ is much greater than the force due to
self-interaction, $g/\rho^2\ll \tilde F$.  This yields the condition $g^3
\sqrt{\tilde\Lambda} /T_2^2 \ll 1$, or
\beq
\left({gM_{(6)}^2\over{T_2}}\right)^2{1\over{n_0}}\ll 1,
\eeq
where in the last step we used Eq.(\ref{n0}).  In a large landscape,
this condition is satisfied, as long as the branes are not too light.

It can be easily verified that the tunneling action
  (\ref{Stestbrane}) with ${\tilde F}$ from (\ref{F}) is smaller than
  the Hawking-Moss action (\ref{SHMcompare}) by a factor of the order
\beq
{1\over{n_0^2}}\left({T_2\over{gM_{(6)}^2}}\right)^4 \ll 1 .
\eeq 
Thus, for light branes, flux tunneling proceeds much more rapidly than
decompactification tunneling.

\subsection{The $4d$ perspective}

Our $6d$ model can be reduced to a purely $4d$ scenario following the
steps we described for the $5d$ case in Sec.~III.  The resulting
action is
\beq
S= \int{d^4 x \sqrt{-g}\left({1\over 2} M_p^2 R^{(4)}-{1\over {48}}
  F_{\mu \nu \alpha \beta} F^{\mu \nu \alpha \beta} -
  \Lambda_{(4)}\right)} - T_2 \int d^{(3)}\Sigma + {{\cal Q}\over{3!}} \int
A_{\sigma \tau \lambda}d^{(3)}
\Sigma^{\sigma \tau \lambda},
\label{6d-4-form-action}
\eeq
where $\sqrt{- g} =  \sqrt{-\tilde g_4}$, $M_P^2 = V_{S^2} M^4_{(6)}$, 
$\Lambda_{(4)} = V_{S^2} {\tilde \Lambda} - 4 \pi M_{(6)}^4$
and $F_{\mu \nu \alpha \beta} = \sqrt{V_{S^2}}{\tilde F}_{\mu \nu \alpha
  \beta}$, $A_{\sigma \tau \lambda} = \sqrt{V_{S^2}}{\tilde A}_{\sigma
  \tau \lambda}$, ${\cal Q} = g/ \sqrt{V_{S^2}}$ and $V_{S^2} = 4 \pi R^2$. 
The $4d$ values of the four-form that correspond to the $6d$ solution
can now be obtained using Eq.~(\ref{Fdual}),

\beq
F^{\mu \nu \delta \gamma} =  {{\epsilon^{\mu \nu \delta
      \gamma}}\over{\sqrt{-g}}} ~\left({{\sqrt{4 \pi} n}\over {2e R}}
\right) = {{\epsilon^{\mu \nu \delta
      \gamma}}\over{\sqrt{-g}}} ~\left({{g n}\over {\sqrt{4 \pi} R}}
\right) = {{\epsilon^{\mu \nu \delta
      \gamma}}\over{\sqrt{-g}}} ~n {\cal Q}~.
\eeq

The situation in this case is somewhat better than
in the $5d$ model since, even though this action disregards the
change of the size of the internal manifold with $n$, we can see that in
the large landscape limit 

\beq
\left({{\Delta R}\over {R}}\right)_{n=n_0} = {3 \over {2 n_0}} ,
\eeq
so we are justified to use this action to compute tunneling 
rates in the neighborhood of $n=n_0$.

\subsection{Another sector of the $6d$ Landscape}

Finally, we should comment on another flux compactification sector of
our $6d$ theory. The existence of this branch of the landscape is more
easily understood in the dual picture, where we have a four-form field
flux that one could turn on a four sphere. One can then find solutions
of this model with two large spacetime dimensions, having de
  Sitter, Minkowski, or anti-deSitter geometry, and with the remaining
  4 dimensions compactified on a $S^4$. We can study tunneling
processes between different values of the {\it monopole-like} number
on the 4-sphere or go to the Maxwell description where the 4-form flux
along the internal dimensions gets dualized to an electric field along
the large spatial dimension.  It is easy to see then that one can
understand the tunneling between vacua in this sector as the Schwinger
decay of this electric field.

One could also ask whether or not there is an instanton that
interpolates between the two sectors in this model. This would have to
be a more complicated instanton than the ones we have been discussing,
as it should involve a topology change to be able to interpolate between
the different compactification schemes. This is an important point,
since the existence of this type of instanton is necessary in order
for the multiverse to explore all the sectors of the landscape.

\section{Bubble collisions}

The structure of bubbles resulting from flux tunneling in our model is
rather unusual.  These bubbles are bounded by codimension-3 branes
which are localized in the extra dimensions.  This has important
implications for bubble collisions.

\begin{figure}
\centering\leavevmode
\epsfysize=5cm \epsfbox{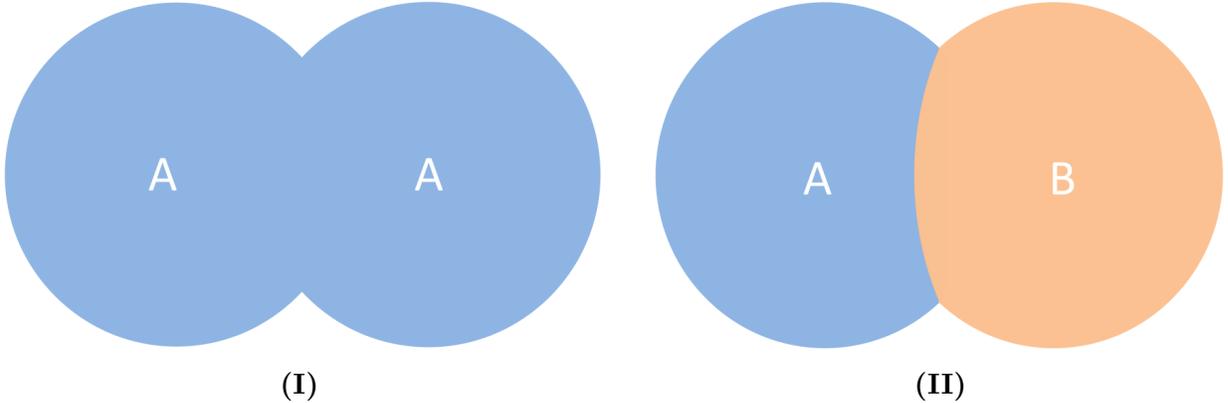}
\put(-372,-10){ {\bf {(I)}}}
\put(-132,-10){ {\bf {(II)}}}
\caption{Two bubbles of type A vacuum merge into each other (I). 
Type A and B vacuum bubbles collide and are separated by a new wall (II).}
\label{Bubblecollisions-1}
\end{figure}

It is usually assumed that when two bubbles of the same vacuum
collide, their domain walls annihilate in the vicinity of the
collision point, with great energy release, and the two bubbles merge
(see Fig.~\ref{Bubblecollisions-1}).  At late times after the collision, the resulting
configuration has the form of two expanding spheres which are joined
along a circle of ever expanding radius.  In the case of bubbles with
different vacua, a similar configuration is formed, but now the
colliding walls merge to produce a new wall that separates the two
vacua inside the bubbles (Fig.~\ref{Bubblecollisions-1}).

\begin{figure}
\centering\leavevmode
\epsfysize=6.5cm \epsfbox{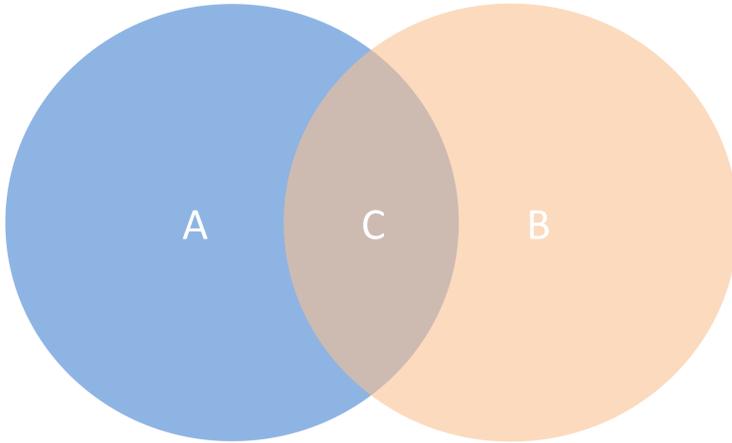}
\caption{Flux vacua of type A and B propagate into one another forming 
a new type C vacuum.}
\label{Bubblecollisions-2}
\end{figure}

In contrast, the branes separating flux vacua in different bubbles are
generally localized at different points in the internal manifold and
will therefore miss one another in the colliding bubbles.  So the
branes will not merge or annihilate, and the bubbles will simply
propagate into one another, forming a new vacuum in the overlap region
(see Fig.~\ref{Bubblecollisions-2}).  For example, if the parent
vacuum has the flux quantum number equal to $n$, and vacua $A$ and 
$B$ both have $n-1$, then vacuum $C$ will have the flux number 
$n-2$. This new type of behavior could have important phenomenological 
consequences for the observable signatures of bubble collisions.

\section{Conclusions}

A generic feature of the multiverse models, inspired by string
theory and inflationary cosmology is the incessant nucleation of
bubbles within bubbles. Thus, in order to understand the multiverse
quantitatively, we have to learn how to calculate bubble
nucleation rates.

In this paper we have set out to study bubble nucleation rates in a 
toy string theory landscape - the $6d$ Einstein-Maxwell model. We have 
shown that vacuum decay can occur via the nucleation of magnetically 
charged 2-branes.  From the $4d$ viewpoint, these branes look like 
expanding bubbles which have their magnetic flux on the inside reduced 
by one unit compared to that on the outside.  We have calculated the 
instanton action for this flux tunneling and compared it to the 
decompactification decay channel. 

We have identified solutions of the Einstein-Maxwell theory
which describe the magnetically charged branes.  They are limiting
cases of the class of solutions previously found by Gregory
\cite{Gregory} and take a particularly simple form in the ``extremal''
case, when the brane tension is simply related to its charge. We
find that for light $(T\ll T_{ext})$ and near-extremal $(T \sim T_{ext})$
branes, flux tunneling proceeds far more rapidly than
decompactification tunneling, while for superheavy branes $(T\gg
T_{ext})$ the two tunneling rates are comparable.

Our model can be easily generalized to include more Maxwell fields
coupled only through gravity. The situation would be very similar to
the one presented here, except that there would be more vacua, and
different types of branes with different charges.

We have also emphasized that the expanding bubbles resulting from
flux tunneling are bounded by higher co-dimension branes, which are
generally localized at different points in the internal dimensions.
We expect, therefore, that in bubble collisions, the branes will
generally miss one another and the bubbles will continue expanding
into each other's interior, forming a new vacuum in the overlap
region.  This may have interesting observational implications, which
we hope to explore in the future.

\section{Acknowledgements}

We would like to thank Ken Olum, Mike Salem and Ben Shlaer for useful
discussions. This work was supported in part by the National Science
Foundation Grants 06533561 (J.J.B-P.) and 0353314 (D.S-P. and A.V.).

\appendix

\section{Higher Dimensional Scalar Compactification}

In this appendix we would like to discuss models of spontaneous
compactification with scalars similar to the ones presented in 
\cite{GellMann}. We will concentrate on simple cases where the
compactification manifold is given by a $q$-sphere and the action is
of the form,

\beq
S=\int{d^{d}{\tilde x} \sqrt{-\tilde g} \left(  {{M_{(d)}^{d-2}}\over 2}
  {\tilde R^{(d)}} - {1\over 2} \partial_{M} {\bf \Phi} \partial^{M}
  {\bf \Phi} - {{\lambda}\over 4} ({\bf \Phi}^2 - \eta^2)^2-{\tilde \Lambda}\right)}~.
\label{general-scalar-d-action}
\eeq
where ${\bf \Phi}$ denotes a vector with $q+1$ elements and we are
mainly interested in the case where the spacetime dimension is $d=
4+q$. It clear that this Lagrangian will have in its spectrum
solitonic solutions (braneworlds similar to the ones discussed in
\cite{global-braneworlds}) of codimension $q+1$ which, as we have
discussed in the main text, will be important for the quantum
tunneling processes we are interested in. Assuming that the scalar
fields remain constrained to the vacuum manifold of the potential, we
can concentrate on the degrees of freedom that parametrize this
manifold.  Thus, we can write the following non-linear sigma model
Lagrangian,
\beq
S=\int{d^{d}{\tilde x} \sqrt{-\tilde g} \left( {{M_{(d)}^{d-2}}\over 2}
  {\tilde R^{(d)}} - {1\over 2} \eta^2
h_{ij}  \partial_{M} \phi^i \partial^{M} \phi^j - {\tilde \Lambda}\right)}
\label{general-nlsm-d-action}
\eeq
where $i,j= 1,...,q$ and $h_{ij}(\phi^k)$ denotes the field space metric on
our target manifold, which in this case is a $q$-sphere.

The equations of motion for this model are:
\beq
{2\over{\sqrt{-\tilde g}}} \partial_M \left(\sqrt{-\tilde g} h_{ik}
\tilde g^{MN}\partial_N
\phi^k\right) - \tilde g^{MN} \partial_M \phi^p \partial_N \phi^q 
\left({{\partial h_{pq}}\over{\partial \phi^i}}\right)=0
\label{nlsm-eom},
\eeq
\beq
{\tilde R}_{AB} - {1\over 2} {\tilde g}_{AB} {\tilde R} =
\kappa^2 T_{AB}.
\eeq
where $\kappa^2=1/M_{d}^{d-2}$ and

\beq
T_{AB} = \eta^2 h_{ij} \left(\partial_A \phi^i \partial_B \phi^j - {1\over 2}
{\tilde g}_{AB} \partial_{M} \phi^i \partial^{M} \phi^j \right)
- {\tilde g}_{AB} {\tilde \Lambda}.
\eeq
We will look for solutions of the form,
\beq
ds^2= {\tilde g}_{MN} dx^M dx^N =  {\tilde g}_{\mu \nu} dx^{\mu}
dx^{\nu} + R^2 d\Omega_q^2
\eeq
where $d\Omega_q^2$ denotes the line element for the internal
spacetime which we will take to be a unit $q$-sphere parametrized by the
angles $\varphi^i$.

This metric is such that we have,
\beq
\tilde G_{\mu \nu} = -\left(3 H^2 + {{q(q-1)}\over {2 R^2}}\right)
\tilde g_{\mu \nu}
\eeq
and 

\beq
\tilde G_{ij} =  - \left(6  H^2 + {{(q-1)(q-2)}\over {2R^2}}\right)
\tilde g_{ij} ~.
\eeq
Finally, we look for the simplest solutions for the scalar 
field equations which describe a trivial mapping
between the extra-dimensional $q$-sphere and the scalar field
manifold, in this case a $q$-sphere as well,
\beq
\phi^i ( \varphi^i) =\varphi^i
\eeq
It is then clear that in our ansatz,

\beq
R^2~h_{ij}(\phi^i) = {\tilde g}_{ij}(\varphi^i)
\eeq
and therefore the equations of motion for the nonlinear sigma model
are trivially satisfied.

On the other hand, this field configuration gives rise to the
following energy-momentum tensor,
\beq
T_{\mu \nu} = - \left({{q \eta^2}\over {2 R^2}} + \tilde
\Lambda\right){\tilde g_{\mu \nu}}
\eeq
and
\beq
T_{ij} = - \left(\tilde \Lambda -{{(2-q) \eta^2}\over {2 R^2}}
\right){\tilde g_{ij}}.
\eeq
So Einstein's equations become,

\beq
3 H^2 + {{q (q-1)}\over {2 R^2}} = \kappa^2 \left({{q\eta^2}\over
{2R^2}} +
\tilde \Lambda\right)
\eeq
and

\beq
6  H^2 + {{(q-1)(q-2)}\over {2R^2}} = \kappa^2 \left(\tilde \Lambda
-{{(2-q) \eta^2}\over 
{2 R^2}}\right), 
\eeq
which can be solved to get,

\beq
R^2 = \left({{q+2}\over 2}\right) \left({{(q-1) - \kappa^2
    \eta^2}\over 
{\kappa^2 \tilde \Lambda}}\right)
\eeq
We notice that there are two branches of solutions, depending on
the sign of $\tilde \Lambda$.

\subsection{The $4d$ perspective}

We would like to understand the properties of this compactification
from the four-dimensional perspective. This can be achieved, following
a similar procedure as before, starting with
Eq.~(\ref{general-nlsm-d-action}) as our higher-dimensional action.

Assuming that the metric is of the form
\beq
ds^2= {\tilde g}_{MN} dx^M dx^N = e^{\alpha \psi(x)/M_P} g_{\mu \nu}
dx^{\mu} dx^{\nu} + e^{\beta \psi(x)/M_P} R^{2}~d\Omega_q^2
\label{6D-metric-2}
\eeq
and taking
\beq
\alpha = - \sqrt{{2 q}\over{q+2}},
\eeq
\beq
\beta = 2 \sqrt{{{2}\over{q(q+2)}}},
\eeq
so that the field $\psi$ is canonically normalized in the 4-dimensional
theory, we arrive at
\beq
S= \int{d^4 x \sqrt{-g}\left({1\over 2} M_P^2 R - {1\over 2}
  \partial_{\mu} \psi \partial^{\mu} \psi - V(\psi)\right)}
\eeq
with
\beq
V(\psi)=M_P^2 \left[\left({{q(\kappa^2 \eta^2- (q-1))}\over {2
R^2}}\right) e^{-\left(\sqrt{{2(q+2)}\over {q}}\right)\psi/M_P} +
\kappa^2 \tilde \Lambda e^{- \left(\sqrt{{2 q}\over {2+q}}\right)
\psi/M_P}\right] ~.
\label{Vpsi}
\eeq

Here, we have defined
\beq
M_P^2 = {{V_{S^q}}\over {\kappa^2}} ~, 
\eeq
where $V_{S^q}$ is the volume of a $q$-sphere of radius $R$.

We can now see that the potential (\ref{Vpsi}) has a minimum at
$\psi=0$ if
\beq
R^2 = \left({{q+2}\over 2}\right) \left({{(q-1) - \kappa^2
    \eta^2}\over 
{\kappa^2 \tilde \Lambda}}\right) ~.
\eeq
This is of course the same solution we found before. Furthermore, we can 
calculate the second derivative of the effective
potential around the minimum at $\psi=0$ to get,

\beq
V''(\psi = 0) = - {{ 4 \kappa^2 \tilde \Lambda} \over {2+q}}
\eeq
which shows that only the models with a negative higher dimensional
cosmological constant $\tilde \Lambda$ are stable. The
$\tilde\Lambda>0$ solution will be unstable to
small perturbations in the size of the extra-dimesional
manifold. This means that, similarly to what happened in the
$5d$ case described in the main text, the values of the 
$4d$ cosmological constant in this case are always negative,
since $V(\psi=0) < 0$ for $\tilde \Lambda < 0$.

\end{document}